\documentclass[prc,showpacs,twocolumn,aps]{revtex4}

\usepackage{bm}
\usepackage{amsmath}
\usepackage{epsfig}
\usepackage{psfig}
\usepackage{graphics}
\usepackage{graphicx}

\def\q{\mbox{\boldmath $q$}}
\def\p{\mbox{\boldmath $p$}}
\def\r{\mbox{\boldmath $r$}}
\def\J{\mbox{\boldmath $J$}}
\def\ss{\mbox{\boldmath $\sigma$}}
\newcommand{\bra}[1]{\left\langle #1\right|}
\newcommand{\ket}[1]{\left| #1\right\rangle}

\begin{document}

\title{Effects of nuclear correlations on the $^{16}$O$(e,e'pN)$ reactions 
to discrete final states}

\author{C.~Barbieri}
  \email{barbieri@triumf.ca}
  \homepage{http://www.triumf.ca/people/barbieri}
  \affiliation
     {TRIUMF, 4004 Wesbrook Mall, Vancouver, 
          British Columbia, Canada V6T 2A3 \\  }

\author{C.~Giusti}
  \email{giusti@pv.infn.it}
   \affiliation
    {Dipartimento di Fisica Nucleare e Teorica dell'Universit\`a degli Studi di
      Pavia\\
     and Istituto Nazionale di Fisica Nucleare, Sezione di Pavia, 
     I-27100 Pavia, Italy}

\author{F.~D.~Pacati}
  \email{pacati@pv.infn.it}
   \affiliation
     {Dipartimento di Fisica Nucleare e Teorica dell'Universit\`a degli Studi 
     di Pavia\\
     and Istituto Nazionale di Fisica Nucleare, Sezione di Pavia, 
     I-27100 Pavia, Italy}
 
\author{W.~H.~Dickhoff}
  \email{wimd@wuphys.wustl.edu}
  \homepage{http://www.physics.wustl.edu/~wimd}
  \affiliation
     {Department of Physics, Washington University,
	 St.Louis, Missouri 63130, USA \\  }

\date{\today}

\begin{abstract}
 Calculations of the $^{16}$O$(e,e'pN)$ cross sections to the ground state and
first excited levels of the $^{14}$C and $^{14}$N nuclei are presented.
 The effects of nuclear fragmentation have been obtained in a
self-consistent approach and are accounted for in the determination
of the two-nucleon removal amplitudes.
 The Hilbert space is partitioned in order to compute the contribution
of both long- and short-range effects in a separate way.
 Both the two-proton and the proton-neutron emission
cross sections have been computed within the same models for the
reaction mechanism and the contribution from nuclear structure, with the
aim of better comparing the differences between the two physical processes.
 The $^{16}$O$(e,e'pp)$ reaction is found to be sensitive to short-range
correlations, in agreement with previous results.
The $^{16}$O$(e,e'pn)$ cross section to $1^+$ final states is dominated
by the $\Delta$ current and tensor correlations.
For both reactions, the interplay between collective (long-range) effects and
short-range and tensor correlations plays an important role.
This suggests that the selectivity of $(e,e'pN)$ reactions to the final state
can be used to probe correlations also beyond short-range effects.
\end{abstract}

\preprint{TRI-PP-04-04}

\maketitle

\section{Introduction}
\label{sec:intro}

Among the various processes that characterize atomic nuclei,
short-range correlations (SRC) play a very important role
in the study of nuclear structure.
It is now understood that the repulsive core of the nuclear
interaction, at small distances, has a decisive influence on the spectral
distribution of nucleons and on the binding properties of both finite
and infinite nuclear systems~\cite{Urbana?,mopara,Wim03,WimBa04}.
%
%
%
%
Photo-induced two-nucleon knockout reactions like $(\gamma,NN)$ and $(e,e'NN)$
appear to be a powerful tool to investigate two-body correlations in 
nuclei. Indeed, the probability that  a real or virtual photon is absorbed
by a pair should be a direct measure of the correlation between the two
nucleons.
The measurements of these cross sections have only become
possible in recent years~\cite{OnAl} by means 
of modern electron beam facilities. 
Studies with a $^{16}$O target have been carried out at the AmPS-facility
at NIKHEF-Amsterdam~\cite{OnAl,Gerco,Ronald} and the MAMI-facility in 
Mainz~\cite{Rosner}.
The comparison of the NIKHEF data with theoretical
calculations~\cite{Ry97,PRC57-pp}
for the $^{16}$O$(e,e'pp)$ reaction has been carried out in
Refs.~\cite{Gerco,Ronald}.
In particular, it has been demonstrated that the transition to the ground
state of $^{14}$C is dominated by the presence of SRC whenever the
two protons are emitted back-to-back with small total momenta.
Therefore, the high experimental cross section observed for this transition
at small missing momenta can be considered a clear signature
of SRC~\cite{Gerco,Ronald}.
Further measurements have been carried out at the MAMI-facility
in Mainz for the $^{16}$O$(e,e'pp)$~\cite{Rosner} reaction and proposed
for the $^{16}$O$(e,e'pn)$ case~\cite{MAMI}.
The resolution achieved in these new experiments allows the separation of
specific excited states in the residual nucleus.

A recent $(e,e'p)$ experiment~\cite{sick97,Daniela}
performed at JLab also was aimed at the direct observations of high-momentum
protons in the nucleus, another clear signature of SRC.
These measurements are expected to produce new
and detailed information on the one-body spectral distribution.
However, due to the high missing energies and momenta required to
observe this consequence of SRC, one is forced to work in a kinematic region 
where the effects
of the final-state interaction tend to overwhelm
the direct signal~\cite{sick97,rescatt}.
The advantage of two-nucleon emission lies in the possibility of ejecting the
correlated pair as a whole, thus seeing the effects of SRC even at small
missing energies and momenta but corresponding to large values of
the relative momentum of the pair. 
On the other hand, several studies~\cite{Ry03,PRC57-pp,PRC60-pn} suggest
that details of the
two-nucleon emission cross sections are sensitive not only to SRC. 
Indeed long-range correlations (LRC), at low energy, and the
reaction mechanism are also important.
Moreover, which of these effects is predominant depends on the particular
choice of the  kinematics and on the final state of the residual nucleus,
in particular its angular momentum and parity. 
The latter quantities therefore act as a 
filter for the study of various reaction processes.
Clearly, while this richness of details complicates the extraction of
information related to SRC, it also identifies two-nucleon emission reactions
as a unique tool to probe different aspects of two-body correlations
in finite systems.

 The model of the reaction mechanism employed in Ref.~\cite{PRC57-pp}
was discussed in Ref.~\cite{GP97}. In this work the excitation process
includes the contribution of the usual one-body terms as well as
those two-body currents which involve the intermediate excitation
of the $\Delta$-isobar.
In the present work the improved treatment of the nuclear currents, given
in Refs.~\cite{ic,gnn,mec} will be employed.
The treatment of the final-state interaction accounts for the distorting
effect of their interaction with the remaining nucleons in terms of
an optical potential.
As in previous works, the  mutual interaction between the two outgoing
nucleons will be neglected here.
This was argued in the past by noting that the pair of protons
will leave the nucleus largely back to back making this type
of final-state interaction less important.
 However, recent perturbative calculations on the $(e,e'pp)$
process~\cite{KM,sch1,sch2}
show that this effect can produce a significant increase of the experimental
yield. Work is in progress to include these contributions
completely~\cite{schFadd}.
This issue remains beyond the scope of the present investigation and
will not be further discussed in the following.

An important element in the calculation of the cross section is the 
two-body overlap (or removal) amplitude, which contains the information
on the correlations between the pair of nucleons inside the system.
 These amplitudes were computed in Ref.~\cite{GeAl-sfO16} for two protons by
partitioning the full Hilbert space to obtain a model space large enough
to account for the most relevant LRC. This is based on the assumption that
the effects of SRC concern high relative momentum states (at high energy) and
which are sufficiently
decoupled from the collective motion at low energy
as not to be influenced by it. The LRC were then obtained
by solving the two-hole dressed random phase approximation (hh-DRPA) inside
the  model space, while the distortion due to SRC was included by adding 
appropriate defect functions, computed for the specifically excluded space.
In doing so, the non-locality of the Pauli operator was neglected, resulting
in a set of only few defect functions, essentially independent of the 
center-of-mass (c.m.) motion of the pair.
The resulting two-nucleon spectral function was then employed in the 
calculation of
the $^{16}$O$(e,e'pp)$ cross section in Ref.~\cite{PRC57-pp}.
A similar approach was followed in Ref.~\cite{PRC60-pn} for the 
$^{16}$O$(e,e'pn)$ case also by employing the same model~\cite{GP97,ic,gnn}
of the reaction mechanism.
In this work the two-hole spectral function for a proton-neutron pair was 
obtained by employing a coupled-cluster approach. 
The $S_2$ approximation employed in
Ref.~\cite{PRC60-pn} is quite similar to the evaluation of the short-range
part of the
two-body spectral function in terms of a Brueckner $G$ matrix, as employed in
Ref.~\cite{GeAl-sfO16}, but
does not account as well for the effects of LRC.
However, a full set of defect functions, including their dependence on
the c.m. of the pair, is obtained naturally in this approach.
Given the differences between the above calculations, it is interesting
to compare the emission of both a pp and a pn pair
by evaluating them within the
same description of the nuclear structure effects.
Furthermore, the description of nuclear structure effects related to the
description of the fragmentation of the single-particle strength has been
improved by applying a Faddeev technique to the description of the
internal propagators in the nucleon self-energy~\cite{badi01,badi02}.
This development provides an additional incentive to study the
resulting consequences for the description of two-nucleon removal reactions.
In the present work we pursue these aims by employing the hh-DRPA approach of
Refs.~\cite{GeAl-sfO16,PRC57-pp} while improving on the computation of the
defect functions, in order to obtain a description of SRC comparable
to the one of Ref.~\cite{PRC60-pn}.
 We then apply this model to study both the $^{16}$O$(e,e'pp)$ and
$^{16}$O$(e,e'pn)$ reactions.

In Sec.~\ref{sec:reaction} of this paper the essential steps in the 
calculation of the $(e,e'pN)$ cross sections are summarized.
 The calculation of the two-nucleon removal amplitudes, that describe
the correlations, is discussed in Sec.~\ref{sec:structure}. There,
the approach of separating the contributions of long-range (LRC) and
short-range correlations (SRC) introduced in Ref.~\cite{GeAl-hh} is reviewed
and the present calculation of defect functions is described in some detail.
Sec.~\ref{sec:structure_2h} also summarizes the updated results for the
nuclear structure calculation.
The numerical results  of $^{16}$O$(e,e'pp)$ and $^{16}$O$(e,e'pn)$ cross
sections are presented and discussed in Sec.~\ref{sec:results}, 
while conclusions are drawn in Sec.~\ref{sec:concl}.

\section{Reaction mechanism of the (\lowercase{e,e$'$p}N) cross 
sections}
\label{sec:reaction}

The coincidence cross section for the reaction induced by an electron 
with momentum $\p_{0}$ and energy $E_{0}$, with $E_{0}=|\p_{0}|=p_{0}$, where 
two nucleons, with momenta $\p'_{1}$ and $\p'_{2}$ 
and energies $E'_{1}$ and $E'_{2}$, are ejected from a nucleus is given, in the
one-photon exchange approximation and after 
integrating over $E'_{2}$, by~\cite{Ox,GP} 
\begin{equation}
\frac{{\mathrm d}^{8}\sigma}{{\mathrm d}E'_{0}{\mathrm d}\Omega
{\mathrm d}E'_{1}{\mathrm d}\Omega'_{1} 
{\mathrm d}\Omega'_{2}} = K \Omega_{\mathrm f} f_{\mathrm{rec}} 
|j_\mu J^\mu|^2 .
\label{eq:cs}
\end{equation}
In Eq.~(\ref{eq:cs}) $E'_{0}$ is the energy of the scattered electron with
momentum $\p'_{0}$, $K = e^4{p'_{0}}^2/4\pi^2 Q\,^4$ where 
$Q^2 = \q\,^2 - \omega^2$, with $\omega = E_{0} - E'_{0}$ and 
$\q = \p_0 - \p'_0$, is the four-momentum transfer. The quantity
$\Omega_{\mathrm f} = p'_{1} E'_{1} p'_{2} E'_{2}$ is the phase-space 
factor and integration over $E'_{2}$ produces the recoil factor 
\begin{equation}
f_{\mathrm{rec}}^{-1} = 1 - \frac{E'_{2}}{E_{\mathrm B}} \, \frac{\p'_{2}\cdot 
\p_{\mathrm B}}{|\p'_{2}|^2},
\end{equation}
where $E_{\mathrm B}$ and $\p_{\mathrm B}$ are the energy and momentum of the 
residual nucleus. The cross section is given by the square of the scalar
product of the relativistic electron current $j^\mu$ and of the nuclear 
current $J^\mu$, which is given by the Fourier transform of the transition 
matrix elements  of the charge-current density operator between initial and 
final nuclear states
\begin{equation}
J^\mu (\q) = \int \bra{ \Psi_{\mathrm{f}} } \hat{J}^\mu(\r) 
\ket{\Psi_{\mathrm{i}} }
{\mathrm{e}}^{\,{\mathrm{i}}{\bm{q}} \cdot
\bm{r}} {\mathrm d}\r.
\label{eq:jm}
\end{equation} 

If the residual nucleus is left in a discrete eigenstate of its Hamiltonian,
i.e. for an exclusive process, and under the assumption of a direct knockout
mechanism, the matrix elements of Eq.~(\ref{eq:jm})
can be written as~\cite{GP,GP97}
\begin{eqnarray}
& J^{\mu}({\mbox{\boldmath $q$}}) & = \int
\Psi_{\rm{f}}^{*}({\mbox{\boldmath $r$}}_{1}
{\mbox{\boldmath $\sigma$}}_{1},{\mbox{\boldmath $r$}}_{2}
{\mbox{\boldmath $\sigma$}}_{2})
J^{\mu}({\mbox{\boldmath $r$}},{\mbox{\boldmath $r$}}_{1}
{\mbox{\boldmath $\sigma$}}_{1},{\mbox{\boldmath $r$}}_{2}
{\mbox{\boldmath $\sigma$}}_{2})
\nonumber \\
 & & \times \, \Psi_{\rm{i}}
({\mbox{\boldmath $r$}}_{1}{\mbox{\boldmath $\sigma$}}_{1},
{\mbox{\boldmath $r$}}_{2}{\mbox{\boldmath $\sigma$}}_{2})
{\rm{e}}^{{\rm{i}}
\hbox{\footnotesize {\mbox{\boldmath $q$}}}
\cdot
\hbox{\footnotesize {\mbox{\boldmath $r$}}}
} {\rm d}{\mbox{\boldmath $r$}}
{\rm d}{\mbox{\boldmath $r$}}_{1} {\rm d}{\mbox{\boldmath $r$}}_{2}
{\rm d}{\mbox{\boldmath $\sigma$}}_{1} {\rm d}{\mbox{\boldmath $\sigma$}}_{2}
.  ~
 \label{eq:jq}
\end{eqnarray}
Equation~(\ref{eq:jq}) contains three main ingredients: the final-state wave 
function $\psi_{\rm{f}}$, the nuclear current $J^{\mu}$ and the 
two-nucleon overlap integral $\psi_{\rm{i}}$.

The nuclear current operator is the sum of a one-body and a two-body part. The 
one-body part contains the usual charge operator and the convective and spin 
currents. The two-body current is derived from the effective Lagrangian of 
Ref.~\cite{Peccei}, performing a non relativistic reduction of the 
lowest-order Feynman diagrams with one-pion exchange. We therefore 
have currents 
corresponding to the seagull and pion-in-flight diagrams and to the diagrams 
with intermediate $\Delta$-isobar configurations~\cite{gnn}, i.e.
\begin{eqnarray}
& &\J^{(2)}(\r,\r_{1}\ss_{1},\r_{2}\ss_{2})  = 
\J^{\mathrm{sea}}(\r,\r_{1}\ss_{1},\r_{2}\ss_{2}) \nonumber \\
& & \ + \ \J^{\pi}(\r,\r_{1}\ss_{1},\r_{2}\ss_{2}) 
 +  \J^{\Delta}(\r,\r_{1}\ss_{1},\r_{2}\ss_{2}) . \label{eq:nc}
\end{eqnarray}
Details of the nuclear current components and the values of the parameters 
used in the calculations are given in Refs.~\cite{ic,gnn,mec}.

Equation~(\ref{eq:jq}) involves bound and scattering states, $\psi_{\rm{i}}$ and 
$\psi_{\rm{f}}$, which should consistently be obtained from an 
energy--dependent non-Hermitian Feshbach-type Hamiltonian
for the considered final state of the residual nucleus.
They are 
eigenfunctions of this Hamiltonian at negative and positive energy eigenvalues, 
respectively~\cite{GP,Ox}. 
In practice, it is not possible to achieve this consistency and the
treatment of initial and final state correlations proceeds separately
with different approximations.

The final-state wave function $\psi_{\mathrm {f}}$ includes the interaction of 
each one of the two outgoing nucleons with the residual nucleus. The mutual 
interaction between the two outgoing nucleons has been studied in 
Ref.~\cite{KM} in nuclear matter and, more recently, in two-nucleon knockout
from $^{16}$O in Refs.~\cite{sch1,sch2} within a perturbative treatment. 
This contribution is neglected in the calculation of the present paper, which 
is aimed at investigating the effects of a consistent treatment of SRC and LRC
in the initial state $\psi_{\mathrm {i}}$. Therefore, the 
scattering state is here given by the product of two uncoupled single-particle 
distorted wave functions, eigenfunctions of a complex phenomenological optical 
potential~\cite{Nad} which contains a central, a Coulomb and a spin-orbit term. 
The two-nucleon overlap integral $\psi_{\rm{i}}$ contains the information on 
nuclear structure and correlations. These have been obtained using the same
many-body approach for both pp and pn knock out, as described in the next
section.

\section{Structure amplitudes}
\label{sec:structure}

Following  Ref.~\cite{PRC57-pp}, the two-nucleon overlap
integral $\psi_{\rm{i}}$ [see Eq.~(\ref{eq:jq})] has been computed by solving
the hh-DRPA equation for the two-particle Green's function.
 This approach allows to accurately take into account the effects of
LRC that are important at the small missing energies
considered in this work.
 However, the description of the high-momentum components due to SRC
requires a large number of basis states, including configurations up to
100 $\hbar\omega$ in an harmonic oscillator basis~\cite{CALGM},
which is too large for practical applications.

The guiding principle followed in the present calculation was presented
earlier in Ref.\ \cite{GeAl-hh} and attempts to treat LRC and SRC in a separate 
but consistent way.
This is done by splitting the complete Hilbert space into a model 
space ${\cal P}$, large enough to contain the relevant long-range effects,
and a complementary space ${\cal Q}={\bf 1}-{\cal P}$. 
The general formalism of the effective interactions considers a number of
exact eigenstates of the system, $\ket{\Psi_i}$, that diagonalize the 
complete Hamiltonian $\hat{H}=\hat{T}+\hat{V}$ with eigenvalues $E_i$.
 One then seeks for an effective Hamiltonian $\hat{H}_{eff}$ that is
defined in the space ${\cal P}$ and has the same exact eigenvalues
(see for example Ref.~\cite{KuoPR}),
\begin{equation}
 P \hat{H}_{eff} P \, \ket{\Phi_i} ~=~ E_i \ket{\Phi_i}   \; ,
 \label{eq:Heff}
\end{equation}
where $P$ is the projection
operator onto the space ${\cal P}$ and the eigenstates given by
$\ket{\Phi_i} = P \ket{\Psi_i}$.
 The complete wave functions $\ket{\Psi_i}$, that belong to the full Hilbert
space, can be obtained from the latter by means of
\begin{eqnarray}
  \ket{\Psi_i} &=& \left( {\bf 1} + \hat{\cal X} \right)  \ket{\Phi_i} 
\nonumber \\
             &=& \ket{\Phi_i} ~+~  \ket{ {\cal X}_i }
  \label{eq:Xdef}
\end{eqnarray}
where the correlation operator $\hat{\cal X} = Q \hat{\cal X} P$
converts the component inside the model space into the corresponding
part that belongs to the space ${\cal Q}$.
The latter, $\ket{{\cal X}_i}$, are usually referred to 
as ``defect functions''.

In the present case, the nuclear correlations that lie in the space
${\cal Q}$ are
those due to SRC.
For the case of two nucleons in free space the two-body correlations
can be accounted for completely by solving the following equation
for the transition matrix~$\hat{R}$,
\begin{equation}
 \hat{R}(\omega) ~=~ \hat{V} ~+~ 
    \hat{V} \frac{1}{\omega - \hat{T} + i \eta} \hat{R}(\omega) \; ,
  \label{eq:Rmtx}
\end{equation}
where $\hat{V}$ and $\hat{T}$ are the NN potential and the kinetic energy,
respectively, and $\omega$ is the energy of the correlated pair.
A good approximation of the effective interaction, Eq.~(\ref{eq:Heff}),
that takes into account the effects of short-range distortion, is obtained by
replacing the bare NN interaction $\hat{V}$ with the $G$ matrix,
obtained by solving the Bethe-Goldstone equation
\begin{equation}
 \hat{G}(\omega) ~=~ \hat{V} ~+~ 
    \hat{V} \frac{Q}{\omega - Q \hat{T} Q + i \eta} \hat{G}(\omega) \; .
  \label{eq:Gmtx}
\end{equation}
Equation~(\ref{eq:Gmtx}) accounts for the short-range effects in
a way completely analogous to Eq.~(\ref{eq:Rmtx}) except that the
projection operator $Q$ now allows the intermediate propagation of
the two particles only within the space ${\cal Q}$ (therefore excluding the 
correlations within the model space ${\cal P}$).
 The $G$ matrix~(\ref{eq:Gmtx}) plays the role of a transition
matrix within the model space:
\begin{equation}
\hat{G} \ket{\Phi} ~=~ \hat{V} \ket{\Psi}   \; ,
\end{equation}
where $\ket{\Phi}$ represents the two-body wave function within the
space ${\cal P}$ and $|\Psi>$ is the fully correlated one that
takes into account the distortion due to SRC. The latter regularizes
the otherwise large matrix elements of $\hat{V}$ that would be generated
by its repulsive core at small interparticle distances.
 The correlated wave function is obtained in terms of the uncorrelated
one $|\Phi>$ as
\begin{equation}
 \ket{\Psi} ~=~ \ket{\Phi} ~+~ 
   \frac{Q}{\omega - Q \hat{T} Q + i \eta} \hat{G}(\omega) \ket{\Phi} \; ,
  \label{eq:XvsG}
\end{equation}
which generalizes the Lippmann-Schwinger equation for two particles in the
vacuum and gives an expression for the correlation operator $\hat{\cal X}$
of Eq.~(\ref{eq:Xdef}).

It should be noted that the distinction between long-range (inside the
model space)
and short-range correlations (outside the model space) is an artificial
one. However, it is important to treat those contributions consistently
and to avoid any kind of double counting.
This is an important merit of the present approach~\cite{GeAl-hh}.
The solution of the Bethe-Goldstone equation yields the residual
interaction of the nucleons inside the model space as well as the
defect functions needed to obtain the complete wave function,
as in Eq.~(\ref{eq:Xdef}).

\subsection{Two-nucleon overlap inside the space ${\cal P}$}
\label{sec:structure_2h}

The effects of LRC have been determined by performing
a nuclear structure calculation  within the same shell-model space
as employed in
Ref.~\cite{badi02,badi03}, based on  harmonic oscillator single-particle
(sp) states with an oscillator parameter $b =$ 1.76~fm (corresponding to
$\hbar\omega =$ 13.4~MeV). The space ${\cal P}$ was chosen to contain all
the first four major shells (from $0s$ to $1p0f$) plus the $0g_{9/2}$
orbital.
The results of Refs.~\cite{brand88,brand90,brand91,Rij92,badi03},
suggest that this is large enough to properly account
for the relevant low-energy collective states.
 The effective interaction (\hbox{$G$ matrix}) was derived
from the Bonn-C model of the NN potential~$\hat{V}$~\cite{bonnc}.
Equation~(\ref{eq:Gmtx}) was solved according to the method of
Ref.~\cite{CALGM} by first computing the real reaction matrix associated to
$\hat{R}$, Eq.~(\ref{eq:Rmtx}), in momentum space as a reference interaction.
A correction term was then computed to account for the effects of
the Pauli operator, which was treated in angle-averaged approximation.

 The fragmentation of one-nucleon removal strength is described by
the coupling of the fully dressed sp propagator
to both two-particle~(pp), two-hole~(hh) and particle-hole~(ph) excitations
of the nuclear medium~\cite{badi02}. 
The simultaneous inclusion of all these collective modes into the nucleon
self-energy $\Sigma^\star$ is computationally intensive and requires
the solution of a set of 
Faddeev equations for the two-particle--one-hole and
two-hole--one-particle motions~\cite{badi01}.
This self-energy has been used to solve the Dyson equation for the
one-body propagator 
\begin{equation}
        g_{\alpha\beta}(\omega)
=
        g^0_{\alpha\beta}(\omega)
+
        \sum_{\gamma\delta}
        g^0_{\alpha\gamma}(\omega)
        \Sigma^\star_{\gamma\delta}(\omega)
        g^{\phantom{0}}_{\delta\beta}(\omega) 
\label{eq:DysonE}
\end{equation}
to obtain the one-nucleon removal spectroscopic factors for the 
low-energy final states in $^{15}$N~\cite{GeAl-sfO16,badi02}.
In these works, the depletion of filled orbits by SRC
is also incorporated in the shell-model space calculation by 
including the energy
dependence of the $G$ matrix interaction, which yields an energy-dependent
Hartree-Fock term in the self-energy~\cite{GeAl-sfO16}.
 In Ref.~\cite{badi02}, the collective pp (hh) and ph motion
was studied at the level of the dressed RPA approximation by taking into
account the fragmentation of the one-body spectral function. 
The propagator resulting from Eq.~(\ref{eq:DysonE}) was then substituted
back into the calculation of the collective surface modes and in the Faddeev
equations.
This whole procedure was iterated until full self-consistency was obtained.
The resulting description of the sp strength and corresponding two-hole states
therefore represents an improvement of the description of LRC as compared to
the work of Ref.~\cite{GeAl-hh}.
Nevertheless, there are still features of the two-hole spectrum that cannot
yet be described by the present method.

For the particular case of the two-hole motion, one solves the
Bethe-Salpeter equation~\cite{fetwa,AAA} for the two-nucleon
propagator $G^{II}$ within the shell-model space. In the present hh-DRPA
approach this reduces to
\begin{widetext}
\begin{eqnarray}
\lefteqn{G^{II}_{\alpha\beta , \gamma\delta}
                 (t_1, t_2, t_3, t_4)
= i \left[
        g_{\alpha\gamma}(t_1-t_3)
        g_{\beta\delta}(t_2-t_4)
-
        g_{\alpha\delta}(t_1-t_4)
        g_{\beta\gamma}(t_2-t_3)
\right] }
\nonumber \\
& & -
        \int\limits_{-\infty}^{\infty}
        {\rm d} t'_1 {\rm d} t'_2 {\rm d} t'_3 {\rm d} t'_4
       \sum_{\mu\nu , \kappa\lambda}
       \left[
            g_{\alpha\mu}(t_1-t'_1)
            g_{\beta\nu}(t_2-t'_2)
       \right]
            \bra{\mu \nu}   G 
       (t'_1, t'_2, t'_3, t'_4) \ket{\kappa \lambda}
        G^{II}_{\kappa\lambda , \gamma\delta}
        (t'_3, t'_4, t_3, t_4)
\; , ~ 
 \label{eq:BSEpp}
 \end{eqnarray}
\end{widetext}
where $\bra{\mu \nu} G(t'_1, t'_2, t'_3, t'_4) \ket{\kappa \lambda}$ 
denote the elements of the $G$ matrix interaction.
From the Lehmann representation of the two-nucleon propagator~$G^{II}$
one obtains the reduced matrix elements of the two-nucleon removal tensor
operators~\cite{Ed57,BG77,fetwa}
\begin{equation}
X^{i}_{ab J} =
  \langle\Psi^{i,A-2}_{J}||
            (c_{\tilde{\beta}} c_{\tilde{\alpha}})_J|| \Psi^A_0 \rangle
\label{eq:XabJ}
\end{equation}
where the latin subscripts denote the basis
states without the magnetic quantum number,
$a=\{n_\alpha, l_\alpha, j_\alpha\}$, and
$\tilde{\alpha}=\{n_\alpha, l_\alpha, j_\alpha, -m_\alpha\}$ corresponds
to the time reverse of $\alpha$.

In Eq.~(\ref{eq:XabJ}), the quantities $X^{i}_{ab J}$ represent the 
components of the two-nucleon overlap integral of Eq.~(\ref{eq:jq}) in
the basis states of the model space.
 These can be expanded in terms of harmonic oscillator
wave functions and transformed to a representation in terms of the relative and
c.m. motion. 
 For a discrete final state $i$ of the (A-2)-nucleon system,
with angular momentum quantum numbers $JM$, one obtains
%
%
\begin{eqnarray}
& & \Phi_{\rm{i}}({\mbox{\boldmath $r$}}_{1}{\mbox{\boldmath $\sigma$}}_{1},
{\mbox{\boldmath $r$}}_{2}{\mbox{\boldmath $\sigma$}}_{2}) =  \sum_{nlSjNL} \,
c^{i}_{nlSjNL} \,  R_{NL}(R) \, \nonumber \\
& & \times \ \phi_{nl}(r) \left[\Im ^{j}_{lS}
(\Omega_r,{\mbox{\boldmath $\sigma$}}_1,{\mbox{\boldmath $\sigma$}}_2) \,
Y_{L}(\Omega_R)\right]^{JM},
\label{eq:ppover_P}
\end{eqnarray}
where
\begin{equation}
{\mbox{\boldmath $r$}} = {\mbox{\boldmath $r$}}_{1}-
{\mbox{\boldmath $r$}}_{2}, \, \, \, \, \, 
{\mbox{\boldmath $R$}} = \frac{{\mbox{\boldmath $r$}}_{1}+
{\mbox{\boldmath $r$}}_{2}}{2}
\end{equation}
correspond to the relative and c.m. variables in coordinate space. 
Note that we follow the convention
that denotes lower case for relative and upper case for c.m. coordinate
quantum numbers.
The brackets in Eq.~(\ref{eq:ppover_P}) indicate angular momentum coupling 
of the 
angular and spin wave function $\Im$ of relative motion with the spherical 
harmonic of the c.m. coordinate to the total angular momentum
quantum numbers $JM$. The radial wave functions of the c.m. and
relative motion are denoted by  $R_{NL}$ and $\phi_{nl}$, respectively, 
and correspond to harmonic oscillators
with parameters $b/\sqrt{2}$ and $\sqrt{2}b$~\cite{BMo}.
In Eq.~(\ref{eq:ppover_P}), the nuclear structure information
represented by the amplitudes $X^{i}_{ab J}$ has been
included in the coefficients
\begin{eqnarray}
\lefteqn{ c^{i}_{nlSjNL} = } &&
\nonumber \\
&& \sum_{ab \in {\cal P}} \sum_\lambda
        \frac{(-)^{L+\lambda+j+S} }
             {\sqrt{2}}
        (2\lambda + 1) \hat{j} \hat{S} \hat{j}_a \hat{j}_b
 \left\{ \begin{array}{lll}
                       l_a & l_b & \lambda \\
                       s_a & s_b & S \\
                       j_a & j_b & J
        \end{array}  
      \right\}  
\nonumber \\
&& ~ ~ \times        \langle n l N L \lambda|
        n_a l_a n_b l_b \lambda\rangle
        \left\{\begin{array}{lll}
                       L & l & \lambda \\ 
                       S & J & j  \end{array}
        \right\} X^{i}_{ab J} 
\; ,
\label{eq:Cab}
\end{eqnarray}
where the notation
$\hat{j}=\sqrt{2j+1}$ was used
and the factor $1/\sqrt{2}$ has been inserted to be consistent with
the normalization assumed in Eq.~(\ref{eq:jq}).

\begin{table}
\begin{tabular}{lcccccccc}
\hline
\hline
$^{16}$O$(e,e'pp)^{14}$C$_{g.s.}$ & 
    & $~n~$ & $~N~$ & $~\rho~$  &~~& Refs.~\cite{GeAl-hh,PRC57-pp} 
    &~~& This work \\
     & & & & & & & & and Ref.~\cite{badi02} \\
\hline
$^1S_0 ; L = 0$      &  & 0 & 1 & 2 &  & $-$0.416   &  & $-$0.410 \\
                     &  & 1 & 0 & 2 &  &   +0.415   &  &   +0.416 \\
                     &  & 0 & 0 & 0 &  &   +0.057   &  &   +0.039 \\
                     &  & 1 & 1 & 4 &  & $-$0.069   &  & $-$0.073 \\
                     &  & 0 & 2 & 4 &  &   +0.049   &  & $-$0.006 \\
                     &  & 2 & 0 & 4 &  &   +0.050   &  &   +0.113 \\
                     &  & 1 & 2 & 6 &  &   +0.016   &  &   +0.017 \\
                     &  & 2 & 1 & 6 &  & $-$0.017   &  & $-$0.017 \\
$^3P_1 ; L = 1$      &  & 0 & 0 & 2 &  &   +0.507   &  &   +0.513 \\
                     &  & 0 & 1 & 4 &  &   +0.024   &  &   +0.076 \\
                     &  & 1 & 0 & 4 &  & $-$0.025   &  &   +0.019 \\
$^1D_2 ; L = 2$      &  & 0 & 0 & 4 &  &   +0.016   &  &   +0.015 \\
\hline
\hline
\end{tabular}
    \parbox[t]{1.\linewidth}{
      \caption[]{
Two-proton removal amplitudes from $^{16}$O to the ground state of $^{14}$C,
given in terms of a c.m. and relative motion expansion.
The numbers in the left column are based on the Dressed RPA
calculations described in Ref.~\cite{GeAl-hh}, while those on
the right account for the self-consistency in the nuclear self-energy
obtained in Ref.~\cite{badi02}.
 The quantum number $\rho$ corresponds to the total number of harmonic
oscillator quanta of the pair: $\rho = 2n+l+2N+L$ (lower case for relative
and upper case for c.m. motion).
For instance $\rho = 4$ indicates contributions from two holes in the $sd$
shell.
\label{tab:Amplitudes-GvsF}
}    }
\end{table}

The most important amplitudes for the case of the transition to 
the ground state of $^{14}$C are listed in Tab.~\ref{tab:Amplitudes-GvsF}.
 These amplitudes are compared with the numbers in the left column, that refer
to the calculation of Ref.~\cite{GeAl-hh}.
 The inclusion of the self-consistency effects in~\cite{badi02} (right column)
does not substantially
alter the results for these amplitudes, except for a slight enhancement of
the collectivity of the $^1S_0$ contribution.
Accordingly, the $X^{i}_{ab J}$ principal components obtained for the pp
case are essentially the same as those of Ref.~\cite{GeAl-hh}.
In the calculation of Ref.~\cite{badi02}, the spectroscopic factors for the 
removal of {\em one} nucleon from the $p_{1/2}$ and $p_{3/2}$ 
orbital of $^{16}$O  turned out to be
reduced respectively by a factor of 0.72 and 0.76 as compared with the
independent-particle shell model.
This is still about 10\% 
larger than the factor $\sim$0.65 deduced from
the experiments~\cite{Leu94,WimMarco92,GiustiG}.
Given the competing effects of fragmentaton and of the screening of the
nuclear interaction, it is not clear a priori whether a reduction of the
spectroscopic factors will correspondingly reduce 
the two-nucleon emission cross sections.
Therefore, as in previous work~\cite{PRC57-pp}, we decided not
to replace the calculated spectroscopic factors by the experimental ones
in the present calculation.
%

The most relevant  amplitudes $X^{i}_{ab J}$ obtained for the emission
of a pn pair are given in Tab.~\ref{tab:X_pn}, where they are compared
with the analogous quantities from Ref.~\cite{PRC60-pn}. 
The results indicate that
the mixing of the principal hole states is qualitatively similar
in both calculations, although the hh-DRPA approach tends to favor the
$(0p_{1/2})^{-2}$ and $(0p_{3/2},0p_{1/2})^{-1}$ components in the
g.s. and first excited state of ${}^{14}$N, respectively.
The most important difference is that the present calculation
predicts a sizable contribution for the emission of two nucleons from
particle orbitals above the Fermi level. These components were not included
in the approach of Ref.~\cite{PRC60-pn}.
 The sum of the squared amplitudes of Tab.~\ref{tab:X_pn} for the
transition to the $1^+_1$ and $1^+_2$ states is 0.61 and 0.67, respectively,
in the present approach and was 0.58 for both states in the coupled cluster
calculation.
The latter number was imposed in Ref.~\cite{PRC60-pn} by normalizing the 
amplitudes to the available DPRA results for the pp case~\cite{GeAl-sfO16}.
 The present calculation confirms this result for the pp channel but generates
a higher normalization for the pn amplitudes.
 The above features introduced in the nuclear structure calculation
generate important differences between the cross sections of
Ref.~\cite{PRC60-pn} and the results discussed in Sec.~\ref{sec:results_pn}. 
\begin{table}
\begin{tabular}{lcccccccc}
\hline
\hline
$J^\pi$  & & $(0p_{3/2})^{-2}$  & & $(0p_{3/2},0p_{1/2})^{-1}$ & & $(0p_{1/2})^{-2}$   
  & &    $(0d_{5/2},0d_{3/2})^{-1}$ \\
\hline
  ~ \\
 \multicolumn{7}{l}{This work and Ref.~\cite{badi02}:} \\
$1^+_{1}$   & &   0.033   & &   -0.347   & &   0.699  & &   0.067 \\
$1^+_{2}$   & &   0.264   & &   -0.680   & &  -0.323  & &   0.189 \\
 ~ \\
 \multicolumn{7}{l}{Ref.~\cite{PRC60-pn}:} \\
$1^+_{1}$   & &   0.070   & &   -0.455   & &   0.607   \\
$1^+_{2}$   & &   0.271   & &   -0.544   & &  -0.460   \\
\hline
\hline
\end{tabular}
    \parbox[t]{1.\linewidth}{
      \caption[]{
Proton-neutron removal amplitudes $X^{i}_{ab J}$ from $^{16}$O to the
first two states of $^{14}$N.
 The numbers in the upper part of the table refer to the hh-DRPA results
obtained in this work.
 For comparison, we give the analogous results obtained in the 
coupled cluster calculations of Ref.~\cite{PRC60-pn}
(lower part).
 The normalization of the two-hole amplitudes is higher in the present
work than what was assumed in Ref.~\cite{PRC60-pn}.
\label{tab:X_pn}
}    }
\end{table}


\subsection{Calculation of the defect functions}
\label{sec:structure_defect}

In Eq.~(\ref{eq:Cab}), the first sum runs over the sp
states $a$ and $b$ that belong to the space ${\cal P}$.
Thus the expansion in Eq.~(\ref{eq:ppover_P}) is limited to configurations
within this model space of two major shells above and two major shells
below the Fermi level.
The effects of correlations on the overlap integral involving $\Psi_i$ induced
by the degrees of freedom outside the space ${\cal P}$
can be included as in Eq.~(\ref{eq:Xdef})~[see also Eq.~(\ref{eq:XvsG})].
The effects of SRC are due to close encounters of two nucleons,
which mainly depend on the nuclear density and are not
sensitive to the details of the long-range structure.
Therefore we can assume  that these processes are decoupled from each other.
 Since short-range effects involve high-momentum components, they 
pertain to the degrees of freedom in the space ${\cal Q}$ which are
described equivalently well by both the $R$ and $G$ matrices (since they
differ mainly for their behavior inside the space ${\cal P}$).
Therefore we substitute for $\hat{G}$ in Eq.~(\ref{eq:XvsG})
the corresponding contribution generated by the standard Lippmann-Schwinger
equation for $\hat{R}$ [see Eq.~(\ref{eq:Rmtx})]. 
In the present work, we follow this prescription and compute the defect
functions according to
\begin{equation}
  \ket{ {\cal X}_i } ~=~
    Q \; \left\{ \frac{1}{\omega - \hat{T} + i \eta} \hat{R}(\omega) 
          \right\}  \ket{ \Phi_i } \; ,
  \label{eq:Xi}
\end{equation}
where $\ket{\phi_i }$ is given by Eq.~(\ref{eq:ppover_P}) and the operator $Q$
ensures that all the correlations inside ${\cal P}$  (generated by the
term in curly brackets) are removed, thus avoiding any double counting.
The operator $R$ in Eq.~(\ref{eq:Xi}) acts only on the radial part
$\phi_{nl}$ of Eq.~(\ref{eq:ppover_P}) leaving the contributions from
$R_{NL}$ untouched. The operator $Q$ is computed exactly and in general
can mix the quantum numbers of the relative and c.m. motion, however, without
altering the form of the expansion~(\ref{eq:ppover_P}). Thus the two-nucleon 
overlap amplitude $\Psi_i$ appearing in Eq.~(\ref{eq:jq}) can be
written as
%
%
\begin{eqnarray}
& & \Psi_{\rm{i}}({\mbox{\boldmath $r$}}_{1}{\mbox{\boldmath $\sigma$}}_{1},
{\mbox{\boldmath $r$}}_{2}{\mbox{\boldmath $\sigma$}}_{2}) =  \sum_{lSjNL} \,
 \,  R_{NL}(R) \, \Psi^i_{lSjNL}(r) \nonumber \\
& & \times \left[\Im ^{j}_{lS}
(\Omega_r,{\mbox{\boldmath $\sigma$}}_1,{\mbox{\boldmath $\sigma$}}_2) \,
Y_{L}(\Omega_R)\right]^{JM},
\label{eq:ppover_full}
\end{eqnarray}
where the complete radial components
\begin{equation}
\Psi^i_{lSjNL}(r) = \sum_n c^{i}_{nlSjNL} \phi_{nl}(r)
                ~+~ {\cal X}^i_{lSjNL}(r)
\end{equation}
now include both the effects of LRC and SRC.

The defect functions employed in Ref.~\cite{PRC57-pp} were obtained
by solving the Bethe-Goldstone only for specific partial waves in the relative
motion and disregarding the dependence on the c.m. quantum numbers. This
simplification also involves at least an angle-averaging approximation of
the Pauli operator $Q$~\cite{CALGM}. 
The approach followed here to compute exactly the
operator $Q$ in Eq.~(\ref{eq:Xi}) allows to keep track of the
dependence of ${\cal X}^i_{lSjNL}$ on the c.m. degrees of freedom.
Noting that the
present interest concerns the high-momentum components due to SRC, 
it is natural
to consider Eq.~(\ref{eq:Xi}) as an improvement with respect to the approach
of Ref.~\cite{PRC57-pp}.


\section{Results for proton-nucleon knockout cross sections}
\label{sec:results}

In this section numerical results are presented for the cross sections of the  
reactions $^{16}$O$(e,e'pp)^{14}$C and $^{16}$O$(e,e'pn)^{14}$N  to the
lowest-lying discrete states in the residual nucleus that are expected to be 
strongly populated by direct knockout of two nucleons. 
The main aim of this study is to investigate the
role of correlations, that are consistently included in the two-nucleon overlap
amplitudes for the proton-proton and the proton-neutron emission processes. 
Of particular interest is also the comparison with the $(e,e'pp)$ results of  
Ref.~\cite{PRC57-pp}, as the present approach represents 
an improvement, and with the $(e,e'pn)$ results of Ref.~\cite{PRC60-pn}, where 
a different description was used to calculate the proton-nucleon overlap 
amplitudes. 

\subsection{The $^{16}$O$(e,e'pp)^{14}$C reaction}
 
Calculations have been performed for three low-lying positive parity states of
$^{14}$C: the $0^+$ ground state, the $1^+$ state at 11.3 MeV, and the $2^+$ 
state at
7.67 MeV, which corresponds to the two $2^+$ states at 7.01 and 8.32 in
the experimental spectrum~\cite{Ajz1315}. 
These states are of particular interest since they can be separated 
in high-resolution experiments~\cite{OnAl,Gerco,Ronald,Rosner}.

As an example, we have considered the so-called super-parallel 
kinematics~\cite{GP}, where the knocked-out nucleons are detected parallel and 
antiparallel to the transferred momentum $\q$. In this kinematics, for a fixed
value of the energy and momentum transfer it is possible to explore, for 
different values of the kinetic energies of the outgoing nucleons, all possible 
values of the recoil momentum.

The super-parallel kinematics is favored by the fact that only two structure
functions, the longitudinal and transverse ones, contribute to 
the cross section.
These can in principle be separated by a Rosenbluth plot
in a way analogous to the inclusive electron scattering~\cite{GP}.
 This kinematical setting is also favorable from the experimental point
of view. It 
has been realized in a recent $^{16}$O$(e,e'pp)^{14}$C experiment at 
MAMI~\cite{Rosner} and has been proposed for the first experimental study of 
the $^{16}$O$(e,e'pn)^{14}$N reaction~\cite{MAMI}. The choice of the same 
kinematics for proton-proton and proton-neutron emission is of particular 
interest for the comparison of cross sections and reaction mechanisms and for 
the investigation of correlations and of their contributions in the two 
processes.
 
\begin{figure}
\includegraphics[height=18.5cm, width=8.cm]{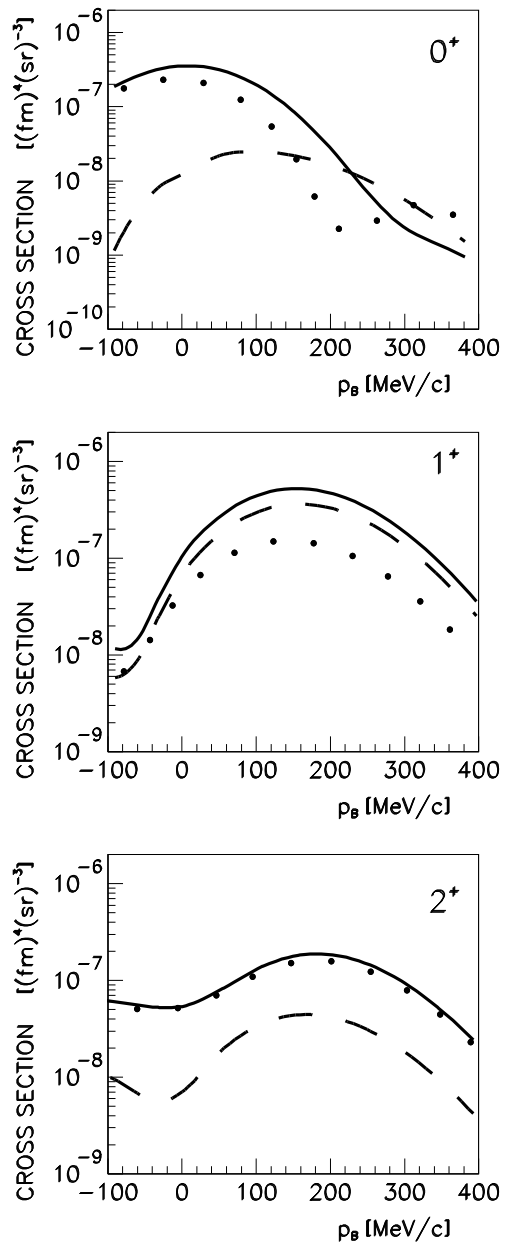}  
\caption[]{
The differential cross section  of the reaction $^{16}$O$(e,e'pp)$ to the
low-lying states of $^{14}$C: the $0^+$ ground state, the  $1^+$ state at 11.31 
MeV, and the  $2^+$ state at 7.67 MeV. A super-parallel kinematics is considered
with $E_{0} = 855$ MeV, $\omega = 215$ MeV $q = 316$ MeV/$c$. Different values 
of the recoil momentum $p_{\mathrm{B}}$ are obtained changing the kinetic 
energies of the outgoing protons. Positive (negative) values of the recoil 
momentum refer to situations where $\p_{\mathrm{B}}$ is parallel (anti-parallel) 
to $\q$. Separate contributions of the one-body and the two-body $\Delta$ 
current are shown by the dotted and dashed lines, respectively. The solid 
curves give the final result.  }
\label{fig:fig1}
\end{figure}
The calculated differential cross sections of the reaction $^{16}$O$(e,e'pp)$ 
to the three final states in super-parallel kinematics are displayed in 
Fig.~\ref{fig:fig1}. The separate contributions of the one-body and the 
two-body $\Delta$ current are also shown in the figure. Note that the seagull 
and pion-in-flight meson-exchange currents do not contribute in proton-proton 
emission, at least in the nonrelativistic limit considered here.  

It was widely discussed in previous studies~\cite{OnAl,PRC57-pp} how resolution
of discrete final states may provide a tool to discriminate between
contributions from one-body currents, due to SRC, and two-body currents. 
The results in Fig.~\ref{fig:fig1} confirm the selectivity of the 
$^{16}$O$(e,e'pp)^{14}$C reaction involving discrete final states which are
differently affected by the two reaction processes.
The one-body current represents the main contribution for the
transitions to the $0^+$ and $2^+$ states, while the transition to the $1^+$
state is dominated by the $\Delta$ current. This result is due to the fact 
that 
the $0^+$  and $2^+$ states are reached predominantly by the removal of $^1S_0$
pairs, whose wave functions are strongly affected by SRC. In contrast, the 
$1^+$ state is reached by the removal of $^3P$ pairs, where SRC only have a
minor effect.

The two-nucleon overlap for each transition is characterized by different
components of relative and c.m. motion. The relative weights of these 
components
determine the weight of the contributions of the one-body and two-body currents
in the cross section and the shape of the recoil-momentum distribution, 
which is driven by the c.m. orbital
angular momentum $L$ of the knocked out pair. This feature is fulfilled 
in a factorized approach~\cite{Got}, where final-state interaction is neglected
and $\p_{\mathrm{B}}$ is opposite to the total momentum of the initial nucleon
pair, and remains valid when final-state
interaction are included~\cite{PRC57-pp,GP97}. 

For the transition to the $0^+$ state there are the following components of 
relative motion: $^1S_0$, which is combined with a c.m. $L=0$, $^3P_1$, 
combined with $L=1$, and $^1D_2$, combined with $L=2$. The contribution from 
$^1D_2$ is negligible and the cross section is dominated by the removal of a 
$^1S_0$ pair, and 
thus by SRC, at low values of $p_{\mathrm{B}}$. The $^3P_1$ component
and thus the $\Delta$ current become more important at larger values of the
momentum, where the contribution of the c.m. component $L=0$ is strongly
reduced. 

The $1^+$ state is predominantly obtained from the $^3P_0$, $^3P_1$, and 
$^3P_2 $ waves, always combined with a $L=1$ c.m. wave function. This explains 
the $p$-wave shape of the recoil-momentum distribution and the dominant role 
of the two-body $\Delta$ current in the cross section. The components $^3P_2$, 
combined with $L=3$, and $^1D_2$, combined with $L=2$,  are also included 
in the calculation, but they give a negligible contribution. 

For the transition to the $2^+$ state there are the following components of
relative motion: $^1S_0$, which is combined with a c.m. $L=2$, $^3P_1$ and 
$^3P_2$, both combined with $L=1$ and $L=3$, and $^1D_2$, combined with $L=0$ 
and $L=2$. 
The main contribution is given by the $^1S_0$ component, which explains the 
$d$-wave shape of the momentum distribution and the dominant role of the 
one-body current in the calculated cross section. 

\begin{figure}
\includegraphics[height=17.4cm, width=8.cm]{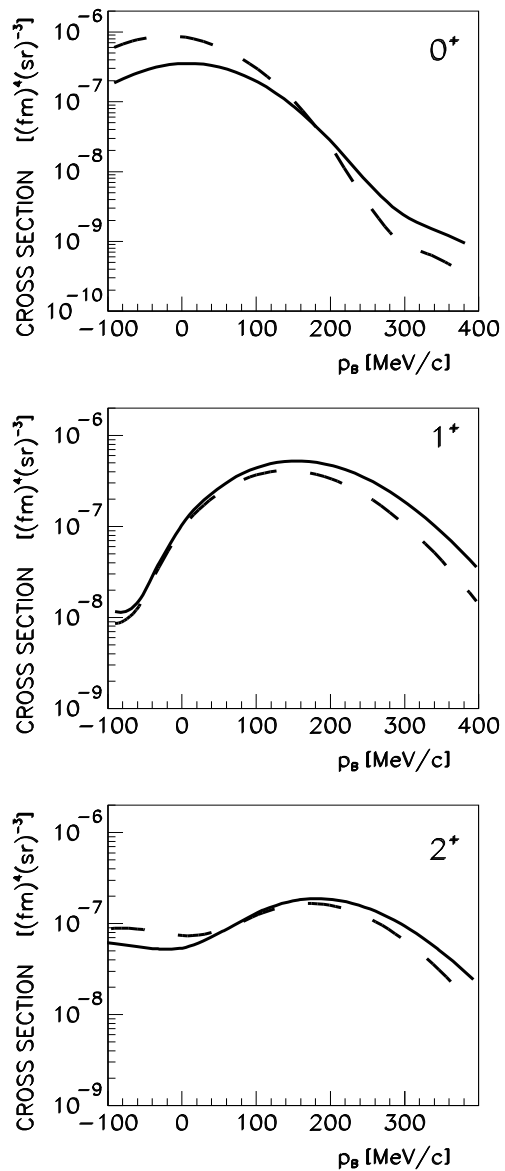}  
\caption[]{
The differential cross section  of the reaction $^{16}$O$(e,e'pp)^{14}$C 
for the same transitions and in the same kinematics as in Fig.~\ref{fig:fig1}. 
The solid lines are the results of the present calculation and the dashed 
lines are the results of Ref.~\cite{PRC57-pp}.
}
\label{fig:fig2}
\end{figure}
These results do not change the qualitative features of the cross section 
calculated in Ref.~\cite{PRC57-pp}. The quantitative 
differences are displayed in Fig.~\ref{fig:fig2}. These differences are 
produced by the detailed treatment of LRC in the removal amplitudes and by the 
new calculation of the defect functions, accounting for SRC, in the present 
approach.
The substantial reduction of the cross section for the $0^+$ state at low 
values
of the recoil momentum is produced by the new defect functions, while the
increase at higher momenta is the result of the combined effect of the new
amplitudes and defects functions. Qualitatively similar but smaller effects 
are found for the $2^+$ state.
Since the transition to the $1^+$ is not very sensitive to SRC, the 
enhancement of this cross section is predominantly due to the new
removal amplitudes. These differ from the ones of Ref.~\cite{PRC57-pp}
by the contribution from the minor $c^{i}_{nlSjNL}$ coefficients
of Eq.~(\ref{eq:Cab}).

\begin{figure}
\includegraphics[height=7.cm, width=7.cm]{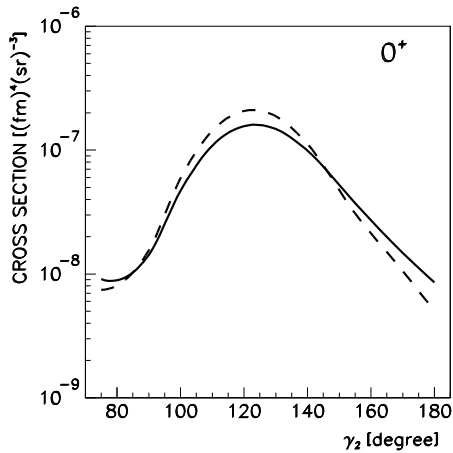}  
\caption[]{
The differential cross section  of the reaction $^{16}$O$(e,e'pp)$ 
to the $0^+$ 
ground state of $^{14}$C as a function of the angle $\gamma_{2}$,  
between $\q$ and $\p'_2$, in a kinematics with $E_0=584$ MeV, 
$\omega=212$ MeV, $q=300$ MeV/$c$, $T'_1=137$ MeV and the angle $\gamma_1$,
between $\p'_1$ and  $\q$, 
$\gamma_1= -30^{\mathrm{o}}$, on the opposite side of the outgoing electron 
with respect to the momentum transfer. Changing the angle $\gamma_2$, 
different values of the recoil momentum $p_{\mathrm{B}}$ are explored in the 
range between $-250$ and $300$ MeV/$c$, including the zero value at 
$\gamma_2 \simeq 120^{\mathrm{o}}$. Line convention as in 
Fig.~\ref{fig:fig2}.}
\label{fig:fig3}
\end{figure}
Another example of the comparison between the results of the present and the
previous approach of Ref.~\cite{PRC57-pp} is shown in  Fig.~\ref{fig:fig3},
for the transition to the  $0^+$ state. Calculations have been performed in a
kinematical setting that was included in the experiments carried out at 
NIKHEF~\cite{Gerco,Ronald}. Also in this kinematics the cross sections are 
dominated by the  removal of a $^1S_0$ pair, and therefore by the one-body 
current, at low momenta, whereas the $^3P_1$ component, and thus the 
$\Delta$ current, becomes important only at larger values of the recoil 
momentum. The differences between the two results are similar to those found 
for the same final state in the super-parallel kinematics of 
Fig.~\ref{fig:fig2}: the cross section calculated in the present approach is 
reduced at low values of $p_{\mathrm{B}}$, where the cross section in 
Fig.~\ref{fig:fig3} has the maximum,  and it is enhanced at higher values of 
$p_{\mathrm{B}}$. Also in this case the reduction 
is produced by the new defect functions. The effect, however, is smaller than 
in  Fig.~\ref{fig:fig2}.

Although the cross sections calculated in the present approach do not 
change the qualitative features of the results obtained in 
Ref.~\cite{PRC57-pp}, the numerical differences confirm that the 
cross sections are very sensitive to the treatment of
correlations in the two-nucleon overlap amplitude. 
SRC, which  are included in the defect functions, predominantly 
affect the part 
involving the one-body current. LRC are accounted for in the removal 
amplitudes 
of Eq.~(\ref{eq:Cab}), which determine the weight of the different components 
of relative and c.m. motion. 
The shape and size of the cross section as well as the role of the one-body 
and two-body currents can thus be affected by both types of correlations. 
Moreover, it should be noted that a consistent treatment of SRC and LRC, 
which represents an important merit of the present approach, entails that the 
two contributions are not independent. 

\subsection{The $^{16}$O$(e,e'pn)^{14}$N reaction}
\label{sec:results_pn}

Calculations have been performed for the two lowest-lying discrete states in 
the residual nucleus $^{14}$N, both with positive parity and $T=0$: 
the 1$^+_1$ ground state and the 1$^+_2$ state at 3.95 MeV. 

\begin{figure}[b]
\includegraphics[height=5.2cm, width=9.5cm]{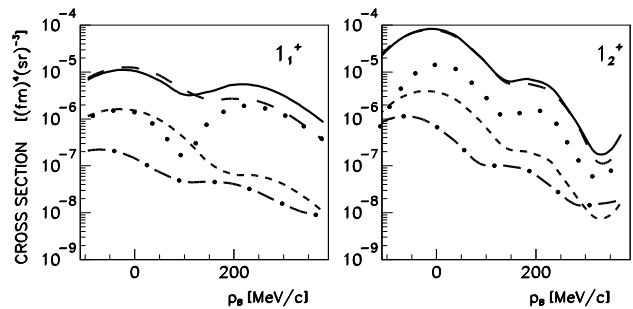}  
\caption[]{
The differential cross section  of the reaction $^{16}$O$(e,e'pn)$ to the 
$1^+_1$ ground state and the $1^+_2$ state (at 3.95 MeV) of $^{14}$N 
in the same super-parallel kinematics as in Fig.~\ref{fig:fig1}. 
The proton is emitted parallel and the neutron antiparallel to the momentum 
transfer.
Separate contributions of the one-body, seagull, pion-in-flight and 
$\Delta$ current are shown by the dotted, short-dashed, dot-dashed and 
long-dashed lines, respectively. The solid line gives the total cross section.}
\label{fig:fig4}
\end{figure}
The differential cross sections of the reaction $^{16}$O$(e,e'pn)^{14}$N
to the two final states in the same
super-parallel kinematics already considered in Fig.~\ref{fig:fig1} for the 
reaction $^{16}$O$(e,e'pp)^{14}$C are displayed in Fig.~\ref{fig:fig4}.
Separate contributions of the different terms of the nuclear current are also 
shown in the figure. For both final states the $\Delta$ current gives the most
important contribution: it is dominant over the whole momentum distribution
shown in the figure for the $1^+_2$ state and for recoil-momentum  values up 
to about 100 MeV/$c$ for the ground state. At higher values of 
$p_{\mathrm{B}}$ 
the contribution of the one-body current becomes for $1^+_1 $comparable 
and therefore competitive with the one of the $\Delta$ current. 
The contributions of the seagull and pion-in-flight terms are very small and 
generally much smaller than the one of the one-body current. 

\begin{figure}
\includegraphics[height=5.2cm, width=9.5cm]{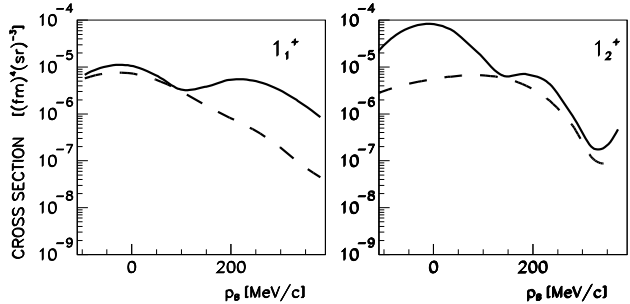}  
\caption[]{
The differential cross section  of the reaction $^{16}$O$(e,e'pn)^{14}$N 
for the same transitions and in the same kinematics as in Fig.~\ref{fig:fig4}. 
The solid lines are the results of the present calculation and the dashed lines 
are the results of Ref.~\cite{PRC60-pn}.}
\label{fig:fig5}
\end{figure}
The comparison with the corresponding cross sections calculated in the approach
of Ref.~\cite{PRC60-pn} is shown in Fig.~\ref{fig:fig5}.
The results of the present approach are always larger than those of 
Ref.~\cite{PRC60-pn}.
For the $1^+_1$ state the differences are within $20\%$ for recoil-momentum
values lower than 100 MeV/$c$ and huge at higher values, where the  
cross section calculated in the present approach overshoots by an order of 
magnitude the result of Ref.~\cite{PRC60-pn}. A different situation is found 
in the $1^+_2 $ state. In this case the present result overshoots by an order 
of magnitude the cross section of Ref.~\cite{PRC60-pn} for values of 
$p_{\mathrm{B}}$ up to $\simeq 100$ MeV/$c$, while  the differences are 
strongly reduced at higher momenta.

Therefore, the two models  produce cross sections which differ both in size 
and shape. Also the contributions of the various terms in the nuclear current
operator can be different in the two calculations. In both cases the 
$\Delta$ current dominates the reaction to the  
$1^+_2 $ state. In contrast, for the  $1^+_1 $ state  the main contribution was
given in Ref.~\cite{PRC60-pn} by the one-body and seagull currents up to 
$p_{\mathrm{B}} \simeq 100$ MeV/$c$ and by the combined effect of these two 
terms with the $\Delta$ current at higher momenta. 

The enhancement of the present cross sections is in part understood
by considering the sum of the squared amplitudes in Tab.~\ref{tab:X_pn},
which in Ref.~\cite{PRC60-pn} were normalized to the hh-DRPA results for
the pp case.
 The difference in the shape of the cross sections should instead
be considered as a result of the different mixing of configurations
in the two cases and the fact that the hh-DRPA description 
considered here allows for pair removal also from particle states. 
Moreover, the inclusion of
fragmentation generates many other coefficients besides those
included in  Tab.~\ref{tab:X_pn}.
We also note that somewhat different models 
are used to calculate the defect functions in the two calculations,
as well as different NN interactions: 
Bonn-C~\cite{bonnc} here and  Argonne $v_{14}$~\cite{v14}
in  Ref.~\cite{PRC60-pn}.

More insight into the cross sections of Fig.~\ref{fig:fig4} and the comparison 
with the results of Ref.~\cite{PRC60-pn} can be obtained from the
separate contributions of the partial waves of relative and c.m. motion which 
are contained in the two-nucleon overlap function. For the transition to the 
two $1^+$ states there are the following relative wave functions: $^3S_1$, 
combined with a c.m. $L= 0$ and $L= 2$, $^1P_1$, combined with $L= 1$, $^3D_1$, 
combined with $L= 0$ and $L= 2$, $^3D_2$ and $^3D_3$, both combined with 
$L= 2$.

\begin{figure}
\includegraphics[height=12.2cm, width=9.5cm]{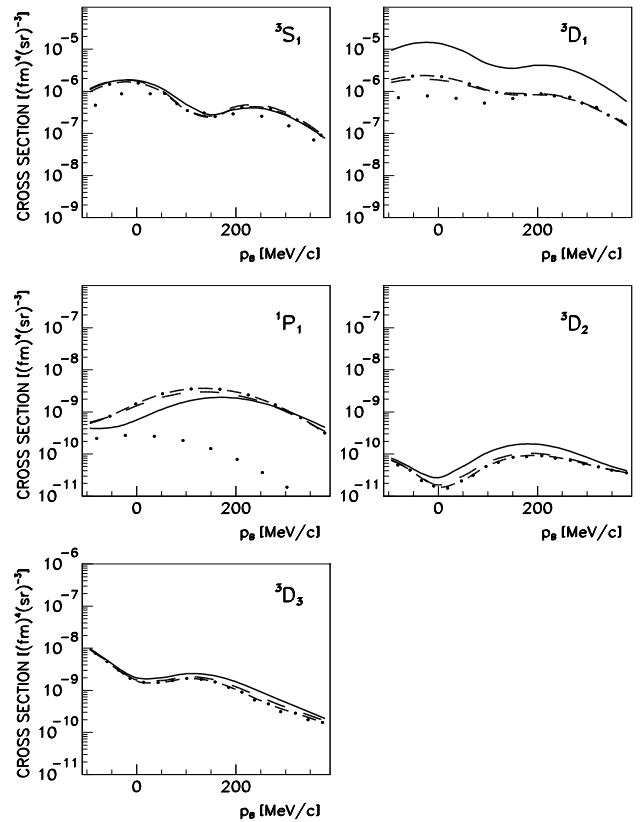}  
\caption[]{
The differential cross section  of the reaction $^{16}$O$(e,e'pn)$ to the 
$1^+_1$ ground state of $^{14}$N in the same kinematics as in 
Fig.~\ref{fig:fig4}. Separate contributions of different partial waves of 
relative motion are drawn: $^3S_1$,  $^3D_1$,  $^1P_1$, $^3D_2$, and $^3D_3$.
The dotted lines give the separate contribution of the one-body current, the 
dot-dashed lines the sum of the one-body and seagull currents, the dashed 
lines the sum of the one-body, seagull and pion-in-flight currents and the 
solid lines the total result, where also the contribution of the 
$\Delta$ current is added.}
\label{fig:fig6}
\end{figure}
The separate contributions of the different partial waves of relative motion 
for the transition to the ground state of $^{14}$N are displayed in 
Fig.~\ref{fig:fig6}. These results can be compared with those shown in Fig.~4 
of Ref.~\cite{PRC60-pn}. Only a very small contribution is obtained from the 
$^1P_1$, $^3D_2$, and $^3D_3$ waves. The $^1P_1$ contribution was practically 
negligible also in the approach of Ref.~\cite{PRC60-pn}, where  the $^3D_2$ 
and $^3D_3$ waves were not included in the calculation. The most important 
contribution is
given in  Fig.~\ref{fig:fig6} by the $^3D_1$ component. This partial wave is  
dominated by the $\Delta$ current, which enhances the cross section by about 
an order of magnitude at low recoil-momentum values. The one-body, and also 
the seagull current, play the main role in  $^3S_1$, but the contribution of 
this partial wave is significant only at large values of the recoil momentum. 
This explains the result in Fig.~\ref{fig:fig5}, where the $\Delta$ current is 
dominant at low momenta and the one-body current is important only above 100 
MeV/$c$. In contrast, in Ref.~\cite{PRC60-pn} the one-body  and the seagull 
current were the main terms at low momenta, where the contribution of $^3S_1$ 
was larger than the one of $^3D_1$. The shape  of the final cross section in 
Fig.~\ref{fig:fig5} is obtained from the combination of the c.m. wave 
functions 
with $L= 0$ and $L= 2$. In the present calculation the $L= 2$ components turn
out to be  more relevant than in Ref.~\cite{PRC60-pn}.

\begin{figure}
\includegraphics[height=5.2cm,width=9.5cm]{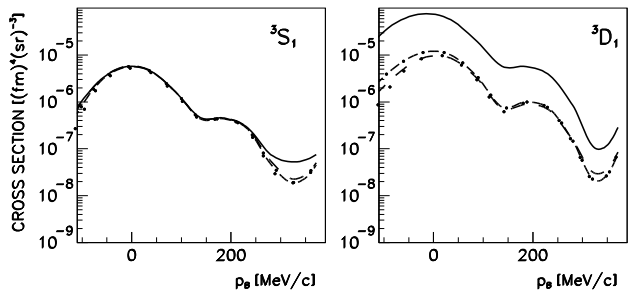}  
\caption[]{
The differential cross section  of the reaction $^{16}$O$(e,e'pn)$ to the 
$1^+_2$ state of $^{14}$N in the same kinematics as in 
Fig.~\ref{fig:fig4}. Separate contributions of  the $^3S_1$ and   $^3D_1$ 
partial waves of relative motion are displayed. Line convention as in 
Fig.~\ref{fig:fig6}.}
\label{fig:fig7}
\end{figure}
For the $1^+_2$ state only the contributions of the most important partial
waves, $^3S_1$ and  $^3D_1$, are drawn in Fig.~\ref{fig:fig7}. The one-body
current is dominant in $^3S_1$ and the $\Delta$ current in $^3D_1$, where it
enhances the cross section by about an order of magnitude. Therefore, the 
final 
cross section is dominated by the $\Delta$ current in the $^3D_1$ component.  
As regards the shape, the contribution of the $L= 0$ wave functions of the c.m.
motion is larger than in Ref.~\cite{PRC60-pn}. 

A crucial contribution to proton-neutron emission is given by tensor
correlations. These correlations, which are mainly due to the strong tensor 
components of the pion-exchange contribution to the NN interaction, are very
important in the wave function of a proton-neutron pair, while they are much
less important for a proton-proton pair. Tensor correlations are accounted 
for in the defect functions and produce correlated wave functions also for 
channels for which the uncorrelated wave function vanishes.  In 
Ref.~\cite{PRC60-pn} the effects of tensor correlations were investigated 
comparing,  for the $^3D_1$ relative wave function, the contribution of the 
components already present in the uncorrelated wave function with the one of 
the components due to the coupling induced by tensor correlations and which are
not present in the uncorrelated wave function. Even if tensor correlations are 
present in all the components, this was a simple way to give an idea of the 
relevance of their contribution. 

\begin{figure}
\includegraphics[height=5.2cm, width=9.5cm]{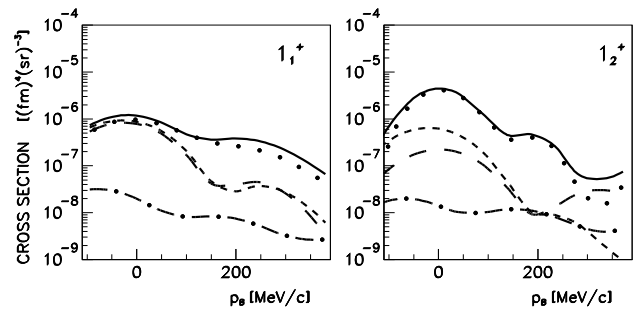}  
\caption[]{
The differential cross section  of the reaction $^{16}$O$(e,e'pn)^{14}$N 
for the same transitions, in the same kinematics and with the same line
convention as in Fig.~\ref{fig:fig4}. 
The defect functions produced by tensor correlations in those channels for 
which the uncorrelated wave function vanishes have been switched off in the 
calculations.}
\label{fig:fig8}
\end{figure}
Likewise here we have performed a calculation of the cross sections neglecting 
the defect functions produced by tensor correlations in those channels for 
which the uncorrelated wave function vanishes. The results for 
the $1^+_1$ and $1^+_2$ states are displayed in Fig.~\ref{fig:fig8}. 
A dramatic reduction of 
the cross section by about one order of magnitude is obtained, for both 
transitions, in comparison with the complete calculations of 
Fig.~\ref{fig:fig4}. This result clear indicates the dominant role of tensor 
correlations in $(e,e'pn)$. The reduction is large for all the terms of the 
nuclear current, but it is huge for the $\Delta$ current, whose contribution 
is reduced by about one order of magnitude in $1^+_1$ and up to about two 
orders 
of magnitude in $1^+_2$. Therefore, the $\Delta$ current, which dominates the 
complete result of Fig.~\ref{fig:fig4}, gives in Fig.~\ref{fig:fig8} a 
contribution comparable to the one of the seagull current and the one-body 
current becomes for both states the most important term in the cross section. 
\begin{figure}
\includegraphics[height=9.5cm,width=9.5cm]{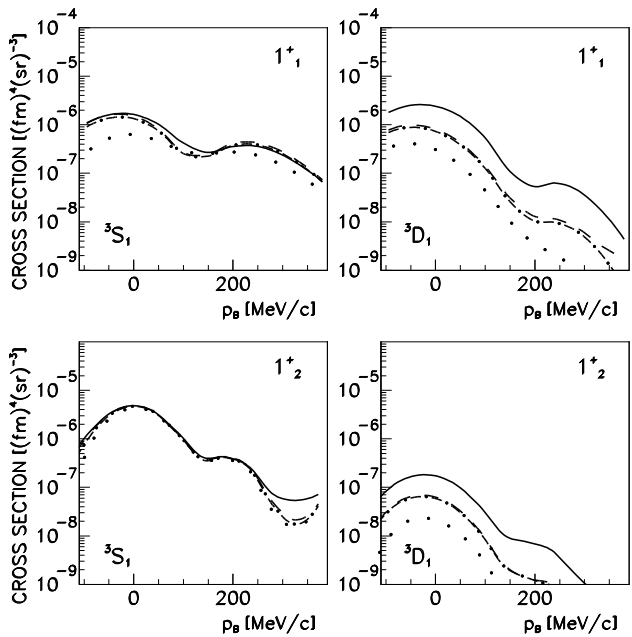}  
\caption[]{
The differential cross section  of the reaction $^{16}$O$(e,e'pn)$ the 
for the same transitions and in the same kinematics as in Fig.~\ref{fig:fig4}. 
The defect functions produced by tensor correlations in those channels for 
which the uncorrelated wave function vanishes have been switched off in the 
calculations. 
Separate contributions of  the $^3S_1$ and   $^3D_1$ partial waves of relative 
motion are displayed. Line convention as in Fig.~\ref{fig:fig6}.}
\label{fig:fig9}
\end{figure}
This result can be seen in more detail in Fig.~\ref{fig:fig9}, where the
separate contributions of the most important partial waves, $^3S_1$ and 
$^3D_1$, are displayed in the calculation where the defect functions 
produced by tensor correlations are neglected. The contribution of $^3S_1$ is 
practically the same as in Figs.~\ref{fig:fig6} and~\ref{fig:fig7}, while the 
contribution of $^3D_1$, is dramatically reduced. The reduction is 
particularly 
strong, of about two orders of magnitude, for the $1^+_2$ state. Therefore, 
in the calculations of  Figs.~\ref{fig:fig8} and~\ref{fig:fig9} the most 
important contribution is given by $^3S_1$ and by the one-body current. 

These results indicate that tensor correlations dominate the $(e,e'pn)$ cross
section. They affect all the terms of the
nuclear current, but produce a particularly strong enhancement of the
$\Delta$- current contribution. This means that also a situation where the 
cross section is dominated by the $\Delta$ current might provide an 
interesting 
and useful tool to investigate tensor correlations. Such a situation can be 
realized in the $(e,e'pn)$ reaction considered here and can be also expected in 
the ($\gamma ,pn)$ reaction, which therefore deserves further 
investigation in the future.
Naturally, the ultimate arbitration of these conjectures must be given by the
experimental data.

\section{conclusions}
\label{sec:concl}

The $^{16}$O$(e,e'pN)$ cross sections have been computed for the transitions
to the ground state, $1^+$ and $2^+$ levels of ${}^{14}$C and to the lowest
two isoscalar $1^+$ states of ${}^{14}$N. 
Both the emissions of a pp or a pn pair have been
computed by employing the same model for the nuclear structure and the
reaction mechanism.

The overlap functions have been computed
by partitioning the Hilbert space, in order to determine the 
contribution of LRC and SRC separately.
 The LRC, describing the collective motion at low energy, are computed within
a model space by solving the hh-DRPA equations and the effects
of fragmentation of the sp strength are taken into account (self-consistently).
The inclusion of SRC is accomplished by determining appropriate
defect functions.
 The present work, improves the treatment of the defect functions 
employed in the $(e,e'pp)$ calculations of Refs.~\cite{GeAl-sfO16,PRC57-pp}
and applies the same many-body approach to the pn emission.

The $^{16}$O$(e,e'pp)$ cross sections are found to be similar to the
results of Ref.~\cite{PRC57-pp}.
 The transitions to the $0^+$ g.s. and the  $2^+$ state of ${}^{14}$C are
shown to be sensitive to the one-body currents and, therefore, to the
effects of SRC. This is in accordance with previous works.
 At small recoil momentum, the reaction rate is found to be lower than
the one of Ref.~\cite{PRC57-pp}. This is due to the different treatment of 
the defect functions employed in this work.
 At high recoil momentum, instead, all the computed $^{16}$O$(e,e'pp)$
cross sections show a slight enhancement due to the interference between the 
LRC amplitudes and the new defect functions.
 However, the main conclusions of previous studies of this reaction are not
changed, including the sensitivity of the effects of correlations
on the choice of the final state.

In contrast, the results for the $^{16}$O$(e,e'pn)$ reaction are found to
deviate from previous calculations~\cite{PRC60-pn}. This is principally due
the different many-body approach employed in this work, which accounts for
the possibility of extracting two nucleons from orbitals above
the Fermi energy (which are partially occupied in the correlated g.s.).
This results in a drastic change of the shape of the cross section.
Moreover, the normalization of the two-hole overlap amplitude is higher in
the hh-DRPA approach for the emission of a pn pair than for a pp pair.
The present calculations suggest that both the transitions
to the $1^+_1$ and $1^+_2$ states of $^{14}$N are dominated by the contribution
from the $\Delta$ current, except for the $1^+_1$ at high recoil momentum where
the one-body current is also important. This situation is different from the
results of Ref.~\cite{PRC60-pn} showing that the reaction rate 
depends sensitively on the details of the LRC as well. 
The present calculation of LRC effects appears to be the most complete
one performed for this specific transition. However, more work on
nuclear structure may be required to check the accuracy of the
results obtained.
 Both the results of this work and of Ref.~\cite{PRC60-pn}, suggest that
the effect of tensor correlations are important for the pn emission (even
dominant in this case) and that they influence the cross
section principally through the $\Delta$ current.
The higher cross section obtained here is a consequence of the interplay
between the details of  LRC and the tensor correlations included in the
defect functions.
This feature could be used to investigate the effects of tensor
correlations by means of $(e,e'pn)$ and ($\gamma ,pn)$ measurements.

In general, all the transitions studied show a strong sensitivity
to the details of nuclear structure and confirm that the importance
of different types of correlations and reaction mechanisms is particular
to the chosen final state.
 While more work can be done on the theoretical side to improve the 
calculation of these cross sections~\cite{badi03,sch1,sch2}, it appears
clear that two-nucleon
emission experiments should be considered as a very powerful tool to probe
various aspects of nuclear correlations, even beyond the SRC.

\acknowledgments
This work was supported in part by the U.S. National Science Foundation
under Grant No.~ PHY-0140316 and in part by the Natural
Sciences and Engineering Research Council of Canada (NSERC).

\appendix



\end{document}